\documentclass[vecphys]{svmult}

\usepackage{makeidx}         
\usepackage{graphicx}        
\usepackage{multicol}        
\usepackage[bottom]{footmisc}

\makeindex             

\def\ha{H$\alpha \ $}

\def\aap{A\&A }
\def\apj{ApJ }

\def\nat{Nature }
\def\araa{ARAA }
\def\mnras{MNRAS }

\begin{document}

\title*{Magnetic Fields in Irregular Galaxies}
\author{Amanda A. Kepley\inst{1}\and
Stefanie M{\"u}hle\inst{2}\and
Eric M. Wilcots\inst{1}\and
John Everett\inst{1,3,4}\and
Ellen Zweibel\inst{1,3}\and
Timothy Robishaw\inst{5}\and
Carl Heiles\inst{5}
}
\institute{ Department of Astronomy, University of Wisconsin--Madison,
  475 N. Charter St., Madison, WI 53706, USA  \texttt{kepley@astro.wisc.edu, ewilcots@astro.wisc.edu, everett@physics.wisc.edu, zweibel@astro.wisc.edu }
\and Department of Astronomy and Astrophysics, University of Toronto,
  50 St. George St., Toronto, Ontario M5S 3H4, Canada \texttt{muehle@astro.utoronto.ca}
\and Center for Magnetic Self-Organization in Laboratory and
  Astrophysical Plasmas
\and Department of Physics, University of Wisconsin--Madison, 1150
  University Ave., Madison, WI 53706-1390, USA
\and 601 Campbell Hall, Department of Astronomy, University of
  California at Berkeley, Berkeley, CA 94720-3411, USA
\texttt{robishaw@astro.berkeley.edu, heiles@astro.berkeley.edu}
}
\authorrunning{Kepley et al.}
%
%
\maketitle

Magnetic fields are an important component of the interstellar
medium. They channel gas flows, accelerate and distribute energy from
cosmic rays, and may be a significant component of the galactic
pressure, especially in low-mass galaxies \cite{beck2004}. A priori,
one wouldn't expect irregular galaxies to have large scale magnetic
fields. The most common dynamo mechanism, the $\alpha-\omega$ dynamo,
relies on differential rotation to stretch small scale fields into
large scale fields \cite{kulsrud_dynamo_review}. Most irregular
galaxies, however, are either solid-body rotators or show little
rotation \cite{grebel_dwarf_review}. Despite this, observations of NGC
4449 \cite{chyzy2000} and the Large Magellanic Cloud (LMC)
\cite{gaensler2005} have revealed the presence of large-scale magnetic
fields in these galaxies.


Previous observations of a small number of irregular galaxies reveal a
range of magnetic field properties. See Table~\ref{tab:gal_properties}
for a summary. Observations of the irregular galaxy NGC 4449
\cite{chyzy2000} show a strong large-scale field. Theoretical work has
suggested that a bar and a fast dynamo are needed to reproduce the
observed field structure \cite{om2000}. The LMC has a weak large-scale
field possibly generated by a cosmic-ray driven dynamo
\cite{gaensler2005}. The magnetic field structures of NGC 6822 and IC
10 \cite{chyzy2003} are weak and almost completely random. The lack of
a large-scale magnetic field in these galaxies may be caused by a
combination of their intense star-formation and their lower rotation
velocities \cite{chyzy2003}.

The goal of this project is to significantly increase the number of
irregular galaxies with observed magnetic field structures to better
answer the following questions: (1) what generates and sustains
large-scale magnetic fields in irregular galaxies? and (2) what causes
the range of observed magnetic field structure?

\section{Measuring Magnetic Fields in  Galaxies}\label{sec:measuring-b-fields}

There are many techniques for measuring magnetic fields \cite{zh97}.
We use diffuse synchrotron emission at centimeter wavelengths as a
tracer of the magnetic field.  High-resolution observations are
crucial for minimizing the effects of beam depolarization. In general,
the optimal observing frequency for these observations is 6\,cm
because there is low Faraday depolarization \cite{burn1966} at this
frequency, but the synchrotron emission is still quite strong. The
increasing strength of synchrotron emission at long wavelengths is
neutralized by an increase in the amount of Faraday depolarization
(which goes as roughly as wavelength squared) at these wavelengths.



We use the Very Large Array (VLA) and single dish radio observations
from either the Green Bank Telescope (GBT) or the Effelsberg 100-m to
obtain radio continuum polarization measurements of several irregular
galaxies at three frequencies: 20\,cm, 6\,cm, and 3\,cm. The VLA
observations have the high resolution necessary for us to detect
small-scale magnetic field structure, while the single dish
observations allow us to correct the VLA observations for unresolved
large-scale structure. Observing at 3 different wavelengths allows us
to separate the free-free emission from the synchrotron emission and
determine rotation measures.

We have selected galaxies that have a range of sizes, rotation rates,
and star formation rates and that complement previous observations
\cite{chyzy2000,gaensler2005,chyzy2003}. See
Table~\ref{tab:gal_properties} for a summary of galaxy properties.

\section{Preliminary Results: NGC 4214 and NGC 1569}\label{sec:results}


NGC 4214 has a slower rotation rate and weaker bar than NGC 4449 and a
higher star formation rate per unit area. From the 6\,cm VLA image of
the continuum emission from this galaxy, we determined the amount of
synchrotron emission using \ha images supplied by Deidre Hunter to
estimate the thermal contribution to the 6\,cm emission. There is very
little polarized emission associated this galaxy. The total field in
this galaxy is about $4 \mu G$ and it is mostly random.


NGC 1569 has the highest star formation rate per unit area in our
sample, and one of the highest star formation rates out of all
galaxies in the local universe. It is possibly ejecting much its
interstellar medium \cite{martin98,lisenfeld04} either through a
pressure-drive, accelerating wind or a detonation.

\begin{figure}
\centering
\includegraphics[angle=-90,width=8.5cm]{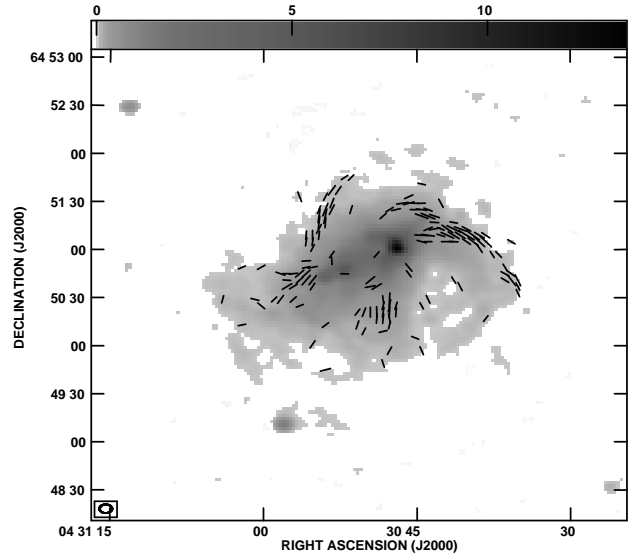}
\caption{Synchrotron emission from NGC 1569 at 6\,cm overlaid with
  polarization vectors. The length of the vector indicates the
  strength of the polarized intensity (1$''$ is 5 $\mu \mathrm{Jy} \
  \mathrm{beam}^{-1}$) and the angle indicates the direction of the
  magnetic field.}
\label{fig:n1569_polplot}       
\end{figure}

Figure~\ref{fig:n1569_polplot} shows a 6\,cm VLA image of the
synchrotron emission from NGC 1569 with vectors showing the
orientation of the magnetic field and the intensity of the polarized
emission. Again, we have used \ha images provided by Deidre Hunter to
estimate the thermal contribution to the 6\,cm emission.

X-ray observations \cite{martin02} suggest that the northern half of
the galaxy is tilted away from the line of sight, so it is not
surprising that we do not see synchrotron emission from this half of
the galaxy. The southern half of the galaxy, however, shows a wealth
of features. An arm of synchrotron emission is seen on the western
edge of the galaxy. Inside this arm, there is an \ha arm and inside
the \ha arm, there is an X-ray arm \cite{martin02}. The large scale
magnetic field seen on the western arm is likely the result of a
compression of gas or a shock. As gas is compressed, it drags the
magnetic field along with it and amplifies the field. Most of the
field lines in this galaxy are roughly perpendicular to the disk,
except at the ends of the disk.

When we compare the 6\,cm image to a 3\,cm image, we see that the
polarization vectors in the southern portion of the galaxy and along
the western arm rotate very little between the two wavelengths. Based
on the observed orientation of the fields between the two wavelengths,
the rotation measure is constrained to be less than 32 $\mathrm{rad} \
\mathrm{m}^{-2}$. This result suggests that the line integral of the
magnetic field along the line of sight times the thermal electron
distribution is small.


The mostly random magnetic field structure of NGC 4214 closely
resembles the magnetic field structure of NGC 6822 and IC 10. This
galaxy probably does not have a strong enough rotation rate or bar to
sustain a large-scale field. NGC 1569 has a large-scale magnetic field
that appears to be shaped primarily by gas outflowing from the disk of
the galaxy. The extent to which the magnetic field shapes or drives
the outflow is the subject of further investigation. We are also
reducing observations of NGC 1313 and NGC 1156 to add their magnetic
field structures to the overall picture of magnetic field structures
in irregular galaxies.


\begin{thebibliography}{10}

\bibitem{beck2004} R.~{Beck}: {Magnetic Fields in the Milky Way and
  Other Spiral Galaxies}. In: {\em How Does the Galaxy Work?}, ASSL
  Vol. 315, ed by E.~J. {Alfaro}, E.~{P{\'e}rez}, and J.~{Franco}
  (Kluwer Academic Publishers, Dordrecht The Netherlands 2004) pp
  277--286

\bibitem{kulsrud_dynamo_review} R.~M. {Kulsrud}: \araa \textbf{37}, 37 (1999)

\bibitem{grebel_dwarf_review} E.~K. {Grebel}: {Dwarf Galaxies in the
  Local Group and in the Local Volume}. In {\em Dwarf galaxies and
  their environment}, ed by K.~S. {de Boer}, R.-J. {Dettmar}, and
  U.~{Klein} (Shaker Verlag, Aachen Germany 2001) pp 45--52

\bibitem{chyzy2000} K.~T. {Chy{\.z}y}, R.~{Beck}, S.~{Kohle},
  et al: \aap \textbf{355}, 128 (2000)

\bibitem{gaensler2005} B.~M. {Gaensler}, M.~{Haverkorn},
  L.~{Staveley-Smith}, et al: Science \textbf{307}, 1610 (2005)

\bibitem{om2000} K.~{Otmianowska-Mazur}, K.~T. {Chy{\.z}y},
M.~{Soida}, and S.~{von Linden}: \aap \textbf{359}, 29 (2000)

\bibitem{chyzy2003} K.~T. {Chy{\.z}y}, J.~{Knapik}, D.~J. {Bomans}, et
  al: \aap \textbf{405}, 513 (2003)

\bibitem{zh97} E.~G. {Zweibel} and C.~{Heiles}: \nat  \textbf{385}, 131 (1997)

\bibitem{burn1966} B.~J. {Burn}: \mnras  \textbf{133}, 67 (1966)

\bibitem{martin98} C.~L. {Martin}: \apj  \textbf{506}, 222 (1998)

\bibitem{lisenfeld04} U.~{Lisenfeld}, T.~W. {Wilding}, G.~G. {Pooley},
  and P.~{Alexander}: \mnras \textbf{349}, 1335 (2004)

\bibitem{martin02} C.~L. {Martin}, H.~A. {Kobulnicky}, and T.~M. {Heckman}: \apj \textbf{574}, 663 (2002)

\end{thebibliography}

\begin{table}
\centering
\caption{Properties of Galaxies in Sample}
\label{tab:gal_properties}       
%
%
\begin{tabular}{lcccccc}
\hline\noalign{\smallskip}
 Galaxy & Optical Extent & $\log (SFR_D)$  & $V_{max}/ \sin (i)$  & Bar? & $B_{uniform}$ & $B_{total}$  \\
        &  kpc & $\mathrm{M_\odot \ yr^{-1} \ kpc^{-2}}$  & $\mathrm{km \ s^{-1}}$ &  & $\mu$G & $\mu$G  \\
\noalign{\smallskip}\hline\noalign{\smallskip}
\multicolumn{7}{c}{Previous Observations \cite{chyzy2000,chyzy2003,gaensler2005}} \\
\noalign{\smallskip}\hline\noalign{\smallskip}
NGC 4449  & $7 \times 5   $    & -2   & 110  &  strong bar    & 6 -- 8     &   14           \\
LMC	  & $9.4 \times 8 $    & -2.9 & 50   &  yes           &  1         & 4.3            \\
NGC 6822  & $2.3 \times 2.0 $ & -1.96 & 52   & yes            &  $< 3$     & $<5$           \\
IC 10     & $2.0 \times 1.7 $ & -1.3  & 30   & no		    &  $<3$      & 5--15          \\
\noalign{\smallskip}\hline\noalign{\smallskip}
\multicolumn{7}{c}{Our Sample} \\
\noalign{\smallskip}\hline\noalign{\smallskip}
NGC 4214  & $10 \times 8 $    & -1.10 & 30 & weak bar       &   $\sim 0$ & 4  \\
NGC 1569  & $2.1 \times 1.0$ & 0.11  & 40 & yes	            &  3--5      & 6  \\
NGC 1156  & $7.5 \times 6 $   & -0.87 & 75 & yes            & ?          & ?  \\
NGC 1313  & $10.6 \times 8.0$ & -0.78 & 89 & strongest bar  & ?          & ?  \\
\noalign{\smallskip}\hline
\end{tabular}
\end{table}

\printindex
\end{document}